\documentclass{article}
\usepackage{spconf,amsmath,graphicx}
\include{pythonlisting}
\usepackage{algorithm}
\usepackage[noend]{algpseudocode}
\usepackage{slashbox}
\usepackage{multirow}
\usepackage{url}
\usepackage{cite}
\usepackage{subcaption}
\makeatletter

\makeatletter
\newcommand{\HEADER}[1]{\ALC@it\underline{\textsc{#1}}\begin{ALC@g}}
\newcommand{\ENDHEADER}{\end{ALC@g}}
\makeatother

\title{Speaker Recognition from raw waveform with SincNet}

\name{Mirco Ravanelli, Yoshua Bengio$^*$}

\address{Mila, Universit\'e de Montr\'eal, $^*$CIFAR Fellow} 
\begin{document}
%
\maketitle
\begin{abstract}
Deep learning is progressively gaining popularity as a viable alternative to i-vectors for speaker recognition. 
Promising results have been recently obtained with Convolutional Neural Networks (CNNs) when fed by raw speech samples directly. Rather than employing standard hand-crafted features, the latter CNNs learn low-level speech representations from waveforms, potentially allowing the network to better capture important narrow-band speaker characteristics such as pitch and formants. Proper design of the neural network is crucial to achieve this goal. 

This paper proposes a novel CNN architecture, called \textit{SincNet}, that encourages the first convolutional layer to discover more meaningful filters. SincNet is based on parametrized sinc functions, which implement band-pass filters. In contrast to standard CNNs, that learn all elements of each filter, only low and high cutoff frequencies are directly learned from data with the proposed method. This offers a very compact and efficient way to derive a customized filter bank specifically tuned for the desired application. 

Our experiments, conducted on both speaker identification and speaker verification tasks, show that the proposed architecture converges faster and performs better than a standard CNN on raw waveforms. 
\end{abstract}
\section{Introduction}
Speaker recognition is a very active research area with notable applications in various fields such as biometric authentication, forensics, security, speech recognition, and speaker diarization, which has contributed to steady interest towards this discipline \cite{Beigi}. Most state-of-the-art solutions are based on the i-vector representation of speech segments \cite{ivector}, which contributed to significant improvements over previous Gaussian Mixture Model-Universal  Background  Models (GMM-UBMs) \cite{gmm_ubm}. 
Deep learning has shown remarkable success in numerous speech tasks \cite{Goodfellow-et-al-2016-Book,lideng,IEEEexample:intro1,ravanelli_thesis,ravanelli_icassp}, including recent studies in speaker recognition \cite{speaker_rec_dnn,dnn_speaker_rec2}.
Deep Neural Networks (DNNs) have been used within the i-vector framework to compute Baum-Welch statistics \cite{Kenny_deepneural}, or for frame-level feature extraction \cite{bottleneck_spk_rec}.
DNNs have also been proposed for direct discriminative speaker classification, as witnessed by the recent literature on this topic \cite{dnn_spk_rec_class2,dnn_spk_rec_class1,dnn_spk_rec_class3,xvector}. 
Most of past attempts, however, employed hand-crafted features such as FBANK and MFCC coefficients \cite{dnn_spk_rec_class2,dnn_speaker_rec_plp,spk_id_mfcc}. These engineered features are originally designed from perceptual evidence and there are no guarantees that such representations are optimal for all speech-related tasks. Standard features, for instance, smooth the speech spectrum, possibly hindering the extraction of crucial narrow-band speaker characteristics such as pitch and formants. To mitigate this drawback, some recent works have proposed directly feeding the network with spectrogram bins \cite{e2e_spk_id,spk_rec_time_freq,voxceleb} or even with raw waveforms \cite{palaz_raw,tara_raw,google_rawmulti,joint7,tuske,dnn_emotion,wavenet,sample_rnn,acoustic_raw_povey,spoofing_raw,raw_speaker_id,verification_raw_ICASSP2018,verification_raw_IS2018}. CNNs are the most popular architecture for processing raw speech samples, since weight sharing, local filters, and pooling help discover robust and invariant representations. 

We believe that one of the most critical part of current waveform-based CNNs is the first convolutional layer. This layer not only deals with high-dimensional inputs, but is also more affected by vanishing gradient problems, especially when employing very deep architectures. The filters learned by the CNN often take noisy and incongruous multi-band shapes, especially when few training samples are available. These filters certainly make some sense for the neural network, but do not appeal to human intuition, nor appear to lead to an efficient representation of the speech signal. 

To help the CNNs discover more meaningful filters in the input layer, this paper proposes to add some constraints on their shape. Compared to standard CNNs, where the filter-bank characteristics depend on several parameters (each element of the filter vector is directly learned), the SincNet convolves the waveform with a set of parametrized sinc functions that implement band-pass filters. The low and high cutoff frequencies are the only parameters of the filter learned from data. This solution still offers considerable flexibility, but forces the network to focus on high-level tunable parameters with broad impact on the shape and bandwidth of the resulting filter.

Our experiments are carried out under challenging but realistic conditions, characterized by minimal training data (i.e., 12-15 seconds for each speaker) and short test sentences (lasting from 2 to 6 seconds). Results achieved on a variety of datasets, show that the proposed SincNet converges faster and achieves better end task performance than a more standard CNN. Under the considered experimental setting, our architecture also outperforms a more traditional speaker recognition system based on i-vectors.

The remainder of the paper is organized as follows. The SincNet architecture is described in Sec.~\ref{sec:sinc}. Sec. \ref{sec:rel_work} discusses the relation to prior work. The experimental setup and results are outlined in Sec.~\ref{sec:setup} and Sec.~\ref{sec:exp} respectively. Finally, Sec.~\ref{sec:conc} discusses our conclusions.

\label{sec:intro}


\section{The SincNet Architecture}
\label{sec:sinc}
\begin{figure}[t!]
 \centering
   \includegraphics[scale=0.80]{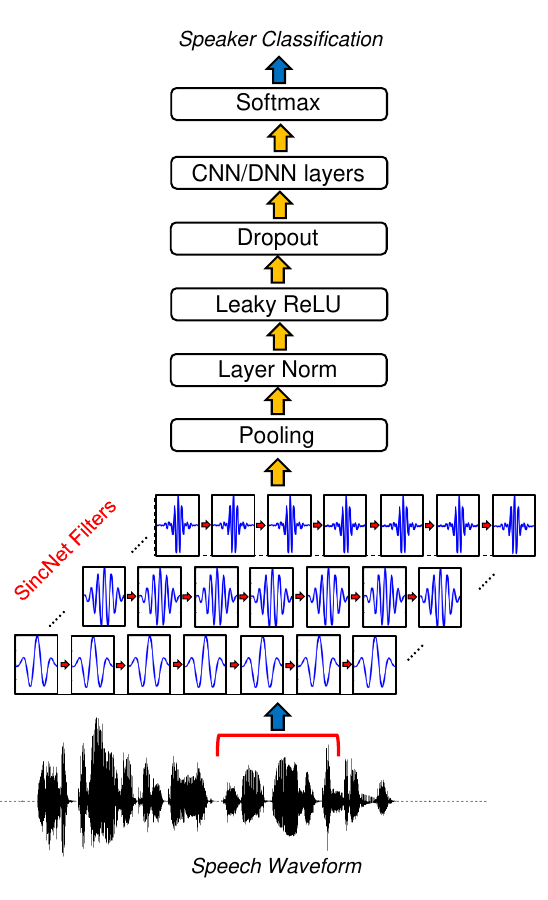}
 \caption{Architecture of SincNet.}
 \label{fig:sinc_arch}
 \end{figure}

The first layer of a standard CNN performs a set of time-domain convolutions between the input waveform and some Finite Impulse Response (FIR) filters \cite{rabiner11}. Each convolution is defined as follows\footnote{Most deep learning toolkits actually compute \textit{correlation} rather than \textit{convolution}. The obtained flipped (mirrored) filters do not affect the results.}:
\begin{equation}
y[n]=x[n]*h[n] = \sum\limits_{l=0}^{L-1} x[l]\cdot h[n-l] 
\end{equation}
where $x[n]$ is a chunk of the speech signal, $h[n]$ is the filter of length $L$, and $y[n]$ is the filtered output. In standard CNNs, all the L elements (taps) of each filter are learned from data. Conversely, the proposed SincNet (depicted in Fig. \ref{fig:sinc_arch}) performs the convolution with a predefined function $g$ that depends on few learnable parameters $\theta$ only, as highlighted in the following equation:

\begin{equation}
y[n]=x[n]*g[n,\theta] 
\end{equation}

A reasonable choice, inspired by standard filtering in digital signal processing, is to define $g$ such that a filter-bank composed of rectangular bandpass filters is employed. In the frequency domain, the magnitude of a generic bandpass filter can be written as the difference between two low-pass filters:

\begin{equation}
G[f,f_1,f_2]= rect\Big(\frac{f}{2f_{2}}\Big) - rect\Big(\frac{f}{2f_{1}}\Big),
\end{equation}
where $f_{1}$ and $f_{2}$ are the learned low and high cutoff frequencies, and $rect(\cdot)$ is the rectangular function in the magnitude frequency domain\footnote{The phase of the $rect(\cdot)$ function is considered to be linear.}.
After returning to the time domain (using the inverse Fourier transform \cite{rabiner11}), the reference function $g$ becomes:

\begin{equation}
g[n,f_1,f_2]= 2f_{2}sinc(2\pi f_2n) - 2f_{1}sinc(2\pi f_1n),
\end{equation}
where the sinc function is defined as $sinc(x)=sin(x)/x$. 

The cut-off frequencies can be initialized randomly in the range $[0,f_s/2]$, where $f_s$ represents the sampling frequency of the input signal.  
As an alternative, filters can be initialized with the cutoff frequencies of the mel-scale filter-bank, which has the advantage  of directly allocating more filters in the lower part of the spectrum, where many crucial clues about the speaker identity are located.
To ensure $f_1\geq0$ and $f_2 \geq f_1$, the previous equation is actually fed by the following parameters:

\begin{align} 
&f_1^{abs}=|f_1| \\ 
&f_2^{abs}=f_1+|f_2-f_1|
\end{align}

Note that no bounds have been imposed to force $f_2$ to be smaller than the Nyquist frequency, since we observed that this constraint is naturally fulfilled during training. 
Moreover, the gain of each filter is not learned at this level. This parameter is managed by the subsequent layers, which can easily attribute more or less importance to each filter output.

An ideal bandpass filter (i.e., a filter where the passband is perfectly flat and the attenuation in the stopband is infinite) requires an infinite number of elements $L$. Any truncation of $g$ thus inevitably leads to an approximation of the ideal filter, characterized by ripples in the passband and limited  attenuation  in  the  stopband.   A popular solution to mitigate this issue is windowing \cite{rabiner11}. Windowing is performed by multiplying the truncated function $g$ with a window function $w$, which aims to smooth out the abrupt discontinuities at the  ends  of  $g$:
\begin{equation}
g_{w}[n,f_1,f_2]= g[n,f_1,f_2] \cdot w[n].
\end{equation}
This paper uses the popular Hamming window \cite{mitra}, defined as follows:
\begin{equation}
w[n]= 0.54-0.46 \cdot cos\Big(\frac{2\pi n}{L}\Big).
\end{equation}
The Hamming window is particularly suitable to achieve high frequency selectivity \cite{mitra}. However, results not reported here reveals no significant performance difference when adopting other functions, such as Hann, Blackman and Kaiser windows. Note also that the filters $g$ are symmetric and thus do not introduce any phase distortions. Due to the symmetry, the filters can be computed efficiently by considering one side of the filter and inheriting the results for the other half.

All operations involved in SincNet are fully differentiable and the cutoff frequencies of the filters can be jointly optimized with other CNN parameters using Stochastic Gradient Descent (SGD) or other gradient-based optimization routines. 
As shown in Fig.  \ref{fig:sinc_arch}, a standard CNN pipeline (pooling, normalization, activations, dropout) can be employed after the first sinc-based convolution.
Multiple standard convolutional, fully-connected or recurrent layers \cite{gru2,ravanelli_is17,li_gru,ravanelli_twin} can then be stacked together to finally perform a speaker classification with a softmax classifier. 

\subsection{Model properties}
The proposed SincNet has some remarkable properties:
\begin{itemize}
\item \textbf{Fast Convergence:}
SincNet forces the network to focus only on the filter parameters with major impact on performance. The proposed approach actually implements a natural inductive bias, utilizing knowledge about the filter shape (similar to feature extraction methods generally deployed on this task) while retaining flexibility to adapt to data. This prior knowledge makes learning the filter characteristics much easier, helping SincNet to converge significantly faster to a better solution.

\item \textbf{Few Parameters:} SincNet drastically reduces the number of parameters in the first convolutional layer. 
For instance, if we consider a layer composed of $F$ filters of length $L$, a standard CNN employs $F \cdot L$ parameters, against the $2F$ considered by SincNet. If $F=80$ and $L=100$, we employ 8k parameters for the CNN and only 160 for SincNet. Moreover, if we double the filter length $L$, a standard CNN doubles its parameter count (e.g., we go from 8k to 16k), while SincNet has an unchanged parameter count (only two parameters are employed for each filter, regardless its length $L$). This offers the possibility to derive very selective filters with many taps, without actually adding parameters to the optimization problem. Moreover, the compactness of the SincNet architecture makes it suitable in the few sample regime. 


\item \textbf{Interpretability}: The SincNet feature maps obtained in the first convolutional layer are definitely more interpretable and human-readable than other approaches. The filter bank, in fact, only depends on parameters with a clear physical meaning.  
\end{itemize}

\section{Related Work} \label{sec:rel_work}
Several works have recently explored the use of low-level speech representations to process audio and speech with CNNs. Most prior attempts exploit magnitude spectrogram features \cite{e2e_spk_id,spk_rec_time_freq,voxceleb,tara_asru2013,learn_fbank_const,fbank_par}. Although spectrograms retain more information than standard hand-crafted features, their design still requires careful tuning of some crucial hyper-parameters, such as the duration, overlap, and typology of the frame window, as well as the number of frequency bins. For this reason, a more recent trend is to directly learn from raw waveforms, thus completely avoiding any feature extraction step. 
This approach has shown promise in speech \cite{palaz_raw,tara_raw,google_rawmulti,joint7,tuske}, including emotion tasks \cite{dnn_emotion}, speaker recognition \cite{raw_speaker_id}, spoofing detection \cite{spoofing_raw}, and speech synthesis \cite{wavenet,sample_rnn}.
Similar to SincNet, some previous works have proposed to add constraints on the CNN filters, for instance forcing them to work on specific bands \cite{tara_asru2013,learn_fbank_const}.  
Differently from the proposed approach, the latter works operate on spectrogram features and still learn all the L elements of the CNN filters. An idea related to the proposed method has been recently explored in \cite{fbank_par}, where a set of parameterized Gaussian filters are employed. This approach operates on the spectrogram domain, while SincNet directly considers raw time domain waveform. 
 
To the best of our knowledge,  this study is the first to show the effectiveness of the proposed sinc filters for time-domain audio processing from raw waveforms using convolutional neural networks. Several past works target speech recognition, while  our study specifically considers a speaker recognition application. The compact filters learned by SincNet are particularly suitable for speaker recognition tasks, especially in a realistic scenario characterized by few seconds of training data for each speaker and short sentences for testing.

\begin{figure*}[t!]
\begin{subfigure}{0.50\textwidth}
\includegraphics[scale=0.65,trim={0cm 0cm 0cm 0cm},clip]{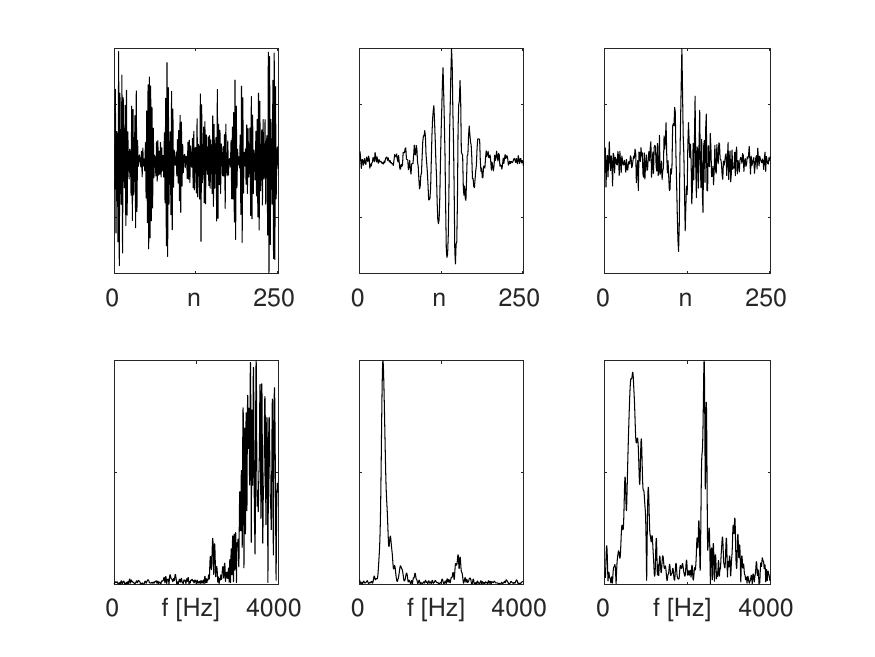}
\caption{CNN Filters}
\label{fig:cnn_filt}
\end{subfigure} \hspace{0.0\textwidth}
\begin{subfigure}{0.50\textwidth}
\includegraphics[scale=0.65]{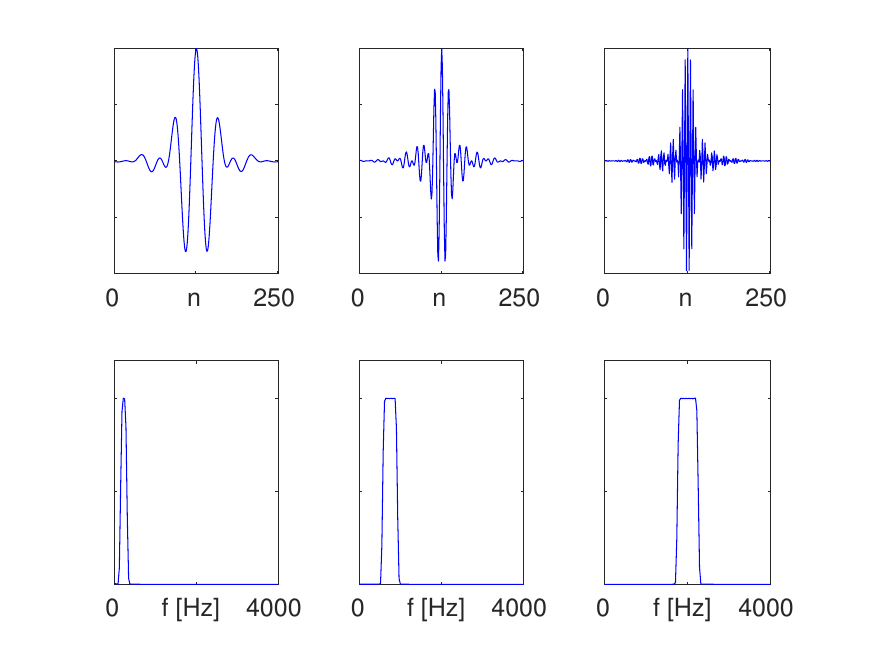}
\caption{SincNet Filters}
\label{fig:sinc_filt}
\end{subfigure}
\caption{Examples of filters learned by a standard CNN and by the proposed SincNet (using the Librispeech corpus). The first row reports the filters in the time domain, while the second one shows their magnitude frequency response.}
\label{fig:ir}
\end{figure*}

\section{Experimental Setup}
The proposed SincNet has been evaluated on different corpora and compared to numerous speaker recognition baselines. 
In the spirit of reproducible research, we perform most experiments using publicly available data such as Librispeech, and release the code of SincNet on GitHub\footnote{\label{foot:code} at \url{https://github.com/mravanelli/SincNet/}.}. In the following sections, an overview of the experimental settings is provided.
\label{sec:setup}
\subsection{Corpora}
To provide experimental evidence on datasets characterized by different numbers of speakers, this paper considers the TIMIT (462 spks, \textit{train} chunk)  \cite{timit} and Librispeech  (2484 spks) \cite{librispeech} corpora. Non-speech intervals at the beginning and end of each sentence were removed. The Librispeech sentences with internal silences lasting more than 125 ms were split into multiple chunks. To address text-independent speaker recognition, the calibration sentences of TIMIT (i.e., the utterances with the same text for all speakers) have been removed. For the latter dataset, five sentences for each speaker were used for training, while the remaining three were used for test. For the Librispeech corpus, the training and test material have been randomly selected to exploit 12-15 seconds of training material for each speaker and test sentences lasting 2-6 seconds.

\subsection{SincNet Setup}
The waveform of each speech sentence was split into chunks of 200 ms (with 10 ms overlap), which were fed into the SincNet architecture. The first layer performs sinc-based convolutions as described in Sec. \ref{sec:sinc}, using 80 filters of length $L=251$ samples. The architecture then employs two standard convolutional layers, both using 60 filters of length 5. Layer normalization \cite{layer_norm} was used for both the input samples and for all convolutional layers (including the SincNet input layer). Next, three fully-connected layers composed of 2048 neurons and normalized with batch normalization \cite{batchnorm} were applied. All hidden layers use leaky-ReLU \cite{leaky_relu} non-linearities. The parameters of the sinc-layer were initialized using mel-scale cutoff frequencies, while the rest of the network was initialized with the well-known ``Glorot" initialization scheme \cite{xavier}. Frame-level speaker classification was obtained by applying a softmax classifier, providing a set of posterior probabilities over the targeted speakers. A sentence-level classification was simply derived by averaging the frame predictions and voting for the speaker which maximizes the average posterior.

Training used the RMSprop optimizer, with a learning rate $lr=0.001$, $\alpha=0.95$, $\epsilon=10^-7$, and minibatches of size 128. 
All the hyper-parameters of the architecture were tuned on TIMIT, then inherited for Librispeech as well. 

The speaker verification system was derived from the speaker-id neural network considering two possible setups. First, we consider the \textit{d-vector} framework \cite{dnn_spk_rec_class2,voxceleb}, which relies on the output of the last hidden layer and computes the cosine  distance between test and the claimed speaker d-vectors. 
As an alternative solution (denoted in the following as \textit{DNN-class}), the speaker verification system can directly take the softmax posterior score corresponding to the claimed identity. The two approaches will be compared in Sec. \ref{sec:exp}.

Ten utterances from impostors were randomly selected for each sentence coming from a genuine speaker. 
Note that to assess our approach on a standard open-set speaker-id task, all the impostors were taken from a speaker pool different from that used for training the speaker-id DNN.

\subsection{Baseline Setups}
We compared SincNet with several alternative systems. 
First, we considered a standard CNN fed by the raw waveform. This network is based on the same architecture as SincNet, but replacing the sinc-based convolution with a standard one. 

A comparison with popular hand-crafted features was also performed. To this end, we computed 39 MFCCs (13 static+$\Delta$+$\Delta\Delta$) and 40 FBANKs using the Kaldi toolkit \cite{kaldi_short}. These features,  computed every 25 ms with 10 ms overlap, were gathered to form a context window of approximately 200 ms (i.e., a context similar to that of the considered waveform-based neural network). A CNN was used for FBANK features, while a Multi-Layer Perceptron (MLP) was used for MFCCs\footnote{CNNs exploit local correlation across features and cannot be effectively used with uncorrelated MFCC features.}. Layer normalization was used for the FBANK network, while batch normalization was employed for the MFCC one. The hyper-parameters of these networks were also tuned using the aforementioned approach. 

For speaker verification experiments, we also considered an i-vector baseline. The i-vector system was implemented with the SIDEKIT toolkit \cite{sidekit}. The GMM-UBM model, the Total Variability (TV) matrix, and the Probabilistic Linear Discriminant Analysis (PLDA) were trained on the Librispeech data (avoiding test and enrollment sentences). GMM-UBM was composed of 2048 Gaussians, and the rank of the TV and PLDA eigenvoice matrix was 400. The enrollment and test phase is conducted on Librispeech using the same set of speech segments used for DNN experiments.

 \begin{figure}[t!]
 \centering
   \includegraphics[scale=0.55]{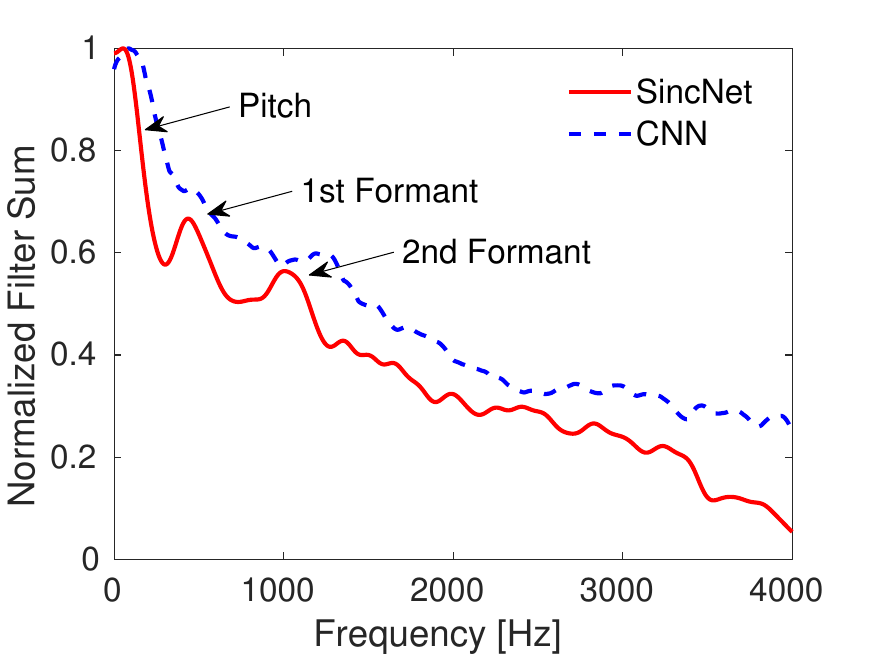}
 \caption{Cumulative frequency response of the SincNet filters.}\label{fig:filt_sum_sum}
 \label{fig:cum}
 \end{figure}
 

\section{Results}
This section reports the experimental validation of the proposed SincNet. First, we perform a comparison between the filters learned by a SincNet and by a standard CNN. We then compare our architecture with other competitive systems on both speaker identification and verification tasks.

\label{sec:exp}

   \begin{figure}[t!]
 \centering
   \includegraphics[scale=0.55]{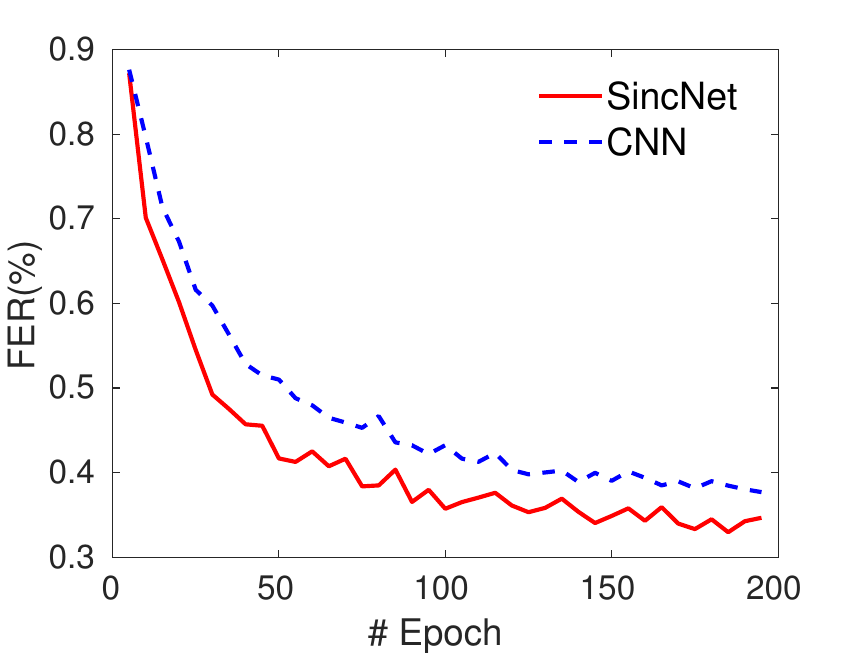}
 \caption{Frame Error Rate (\%) of SincNet and CNN models over various training epochs. Results are reported on TIMIT.}
 \label{fig:conv_curve}
 \end{figure}
 
\subsection{Filter Analysis}
Inspecting the learned filters is a valuable practice that provides insight into what the network is actually learning.
Fig. \ref{fig:ir} shows some examples of filters learned by a standard CNN (Fig. \ref{fig:cnn_filt}) and by the proposed SincNet (Fig. \ref{fig:sinc_filt}) using the Librispeech dataset (the frequency response is plotted between 0 and 4 kHz). As observed in the figures, the standard CNN does not always learn filters with a well-defined frequency response. In some cases the frequency response looks noisy (see the first filter of Fig. \ref{fig:cnn_filt}), while in others assuming  multi-band shapes (see the third filter of the CNN plot). SincNet, instead, is specifically designed to implement rectangular bandpass filters, leading to more meaningful CNN filters. 

Beyond a qualitative inspection, it is important to highlight which frequency bands are covered by the learned filters.
Fig. \ref{fig:cum} shows the cumulative frequency response of the filters learned by SincNet and CNN. Interestingly, there are three main peaks which clearly stand out from the SincNet plot (see the red line in the figure). The first one corresponds to the pitch region (the average pitch is 133 Hz for a male and 234 for a female). The second peak (approximatively located at 500 Hz) mainly captures first formants, whose average value over the various English vowels is indeed 500 Hz. Finally, the third peak (ranging from 900 to 1400 Hz) captures some important second formants, such as the second formant of the vowel $/a/$, which is located on average at 1100 Hz.
This filter-bank configuration indicates that SincNet has successfully adapted its characteristics to address speaker identification. Conversely, the standard CNN does not exhibit such a meaningful pattern: the CNN filters tend to correctly focus on the lower part of the spectrum, but peaks tuned on first and second formants do not clearly appear. As one can observe from Fig. \ref{fig:cum}, the CNN curve stands above the SincNet one. SincNet, in fact, learns filters that are, on average, more selective than CNN ones, possibly better capturing narrow-band speaker clues.


\subsection{Speaker Identification}

Fig. \ref{fig:conv_curve} shows the learning curves of SincNet compared with that of a standard CNN. These results, achieved on the TIMIT dataset,  highlight a faster decrease of the Frame Error Rate ($FER\%$) when SincNet is used. Moreover, SincNet converges to better performance leading to a FER of 33.0\% against a FER of 37.7\% achieved with the CNN baseline. 

\begin{table}[h]
\centering
\begin{tabular}{|l|c|c|c|c|}
\hline
            & TIMIT & LibriSpeech  \\ \hline
DNN-MFCC             &   0.99      &  2.02      \\ \hline
CNN-FBANK       &   0.86      &  1.55     \\ \hline
CNN-Raw       &   1.65     &  1.00       \\ \hline
SINCNET       &   \textbf{0.85}      &  \textbf{0.96}      \\ \hline
\end{tabular}
\caption{Classification Error Rate (CER\%) of speaker identification systems trained on TIMIT (462 spks) and Librispeech (2484 spks) datasets. SincNets outperform the
competing alternatives.}
\label{tab:spk_id_res}
\end{table}

Table \ref{tab:spk_id_res} reports the achieved Classification Error Rates (CER\%). The table shows that  SincNet outperforms other systems on both TIMIT and Librispeech datasets. The gap with a standard CNN fed by raw waveform is particularly large on TIMIT, confirming the effectiveness of SincNet when few training data are available. Although this gap is reduced when LibriSpeech is used, we still observe a 4\% relative improvement that is also obtained with faster convergence (1200 vs 1800 epochs). Standard FBANKs provide results comparable to SincNet only on TIMIT, but are significantly worse than our architecture when using Librispech. With few training data, the network cannot discover filters much better than FBANKs, but with more data a customized filter-bank is learned and exploited to improve the performance.

\subsection{Speaker Verification}
As a last experiment, we extend our validation to speaker verification. Table \ref{tab:spk_ver_res} reports the Equal Error Rate (EER\%) achieved with the Librispeech corpus. All DNN models show promising performance, leading to an EER lower than 1\% in all cases. The table also highlights that SincNet outperforms the other models, showing a relative performance improvement of about 11\% over the standard CNN model.
\textit{DNN-class} models perform significantly better than d-vectors. 
Despite the effectiveness of the later approach, a novel DNN model must be trained (or fine-tuned) for each new speaker added into the pool \cite{raw_speaker_id}. This makes this approach better performing, but less flexible than d-vectors.

\begin{table}[h!]
\centering
\begin{tabular}{|l|c|c|c| }
\hline
            & d-vector & DNN-class \\ \hline
DNN-MFCC   &  0.88  &  0.72    \\ \hline
CNN-FBANK   &  0.60  &  0.37    \\ \hline
CNN-Raw     &  0.58 &  0.36      \\ \hline
SINCNET     &  \textbf{0.51} &  \textbf{0.32}     \\ \hline
\end{tabular}
\caption{Speaker Verification Equal Error Rate (EER\%) on Librispeech datasets over different systems. SincNets outperform the
competing alternatives.}
\label{tab:spk_ver_res}
\end{table}
For the sake of completeness, experiments have also been conducted with standard i-vectors.
Although a detailed comparison with this technology is out of the scope of this paper, it is worth noting that our best i-vector system achieves a EER=1.1\%, rather far from what achieved with DNN systems. It is well-known in the literature that i-vectors provide competitive performance when more training material is used for each speaker and when longer test sentences are employed \cite{i-vect_short,i-vect_short2,i-vect_short3}. Under the challenging conditions faced in this work, neural networks achieve better generalization. 

\section{Conclusions and Future Work}
\label{sec:conc}
This paper proposed SincNet, a neural architecture for directly processing waveform audio. Our model, inspired by the way filtering is conducted in digital signal processing, imposes constraints on the filter shapes through efficient parameterization. SincNet has been extensively evaluated on challenging speaker identification and verification tasks, showing performance benefits for all considered corpora. 

Beyond performance improvements, SincNet also significantly improves convergence speed over a standard CNN, and is more computationally efficient due to exploitation of filter symmetry. Analysis of the SincNet filters reveals that the learned filter-bank is tuned to precisely extract some known important speaker characteristics, such as pitch and formants.
In future work, we would like to evaluate SincNet on other popular speaker recognition tasks, such as VoxCeleb. Although this study targeted speaker recognition only, we believe that the proposed approach defines a general paradigm to process time-series and can be applied in numerous other fields. Our future effort will be thus devoted to extending to other tasks, such as speech recognition, emotion recognition, speech separation, and music processing.

\section*{Acknowledgement}
We would like to thank Gautam Bhattacharya, Kyle Kastner, Titouan Parcollet, Dmitriy Serdyuk, Maurizio Omologo, and Renato De Mori for their helpful comments. This research was enabled in part by support provided by Calcul Qu\'ebec and Compute Canada.



\bibliographystyle{IEEEbib}
\bibliography{strings,refs}

\end{document}